\documentclass[conference]{IEEEtran}
 
\usepackage{cite}

\ifCLASSINFOpdf
\else
\fi
\usepackage{amsmath,amsfonts,amssymb}
\usepackage{mathtools,xparse}
\DeclarePairedDelimiter{\abs}{\lvert}{\rvert}
\DeclarePairedDelimiter{\norm}{\lVert}{\rVert}
\usepackage{url}

\usepackage[]{fancyhdr} %
\newcommand{\changefont}{\fontsize{9}{9}\selectfont}
\ifCLASSOPTIONcompsoc
 \usepackage[caption=false,font=normalsize,labelfont=sf,textfont=sf]{subfig}
\else
 \usepackage[caption=false,font=footnotesize]{subfig}
\fi

\usepackage{multirow}
\usepackage{tabularx}
\usepackage{booktabs}
\usepackage{threeparttable}
\usepackage{siunitx}
\usepackage[table,x11names]{xcolor}
\usepackage{bbm}
\usepackage{balance}

\newcommand*{\myDots}{\ifmmode\mathellipsis\else.\kern-0.13em.\kern-0.13em.\fi} 

\usepackage{afterpage}

\renewcommand{\baselinestretch}{0.97}

\IEEEoverridecommandlockouts
\begin{document}

%


\title{Using Quantile Forecasts for Dynamic Equivalents of Active Distribution Grids under Uncertainty}

\author{\IEEEauthorblockN{Johanna Vorwerk\IEEEauthorrefmark{1}\IEEEauthorrefmark{2},
Thierry Zufferey\IEEEauthorrefmark{1},
Petros Aristidou\IEEEauthorrefmark{3}, and 
Gabriela Hug\IEEEauthorrefmark{1}}
\IEEEauthorblockA{\IEEEauthorrefmark{1}Power Systems Laboratory, ETH Zürich, Zürich, Switzerland}
\IEEEauthorblockA{\IEEEauthorrefmark{3}Sustainable Power Systems Laboratory, Cyprus University of Technology, Limassol, Cyprus}
\IEEEauthorblockA{\IEEEauthorrefmark{2}vorwerkj@ethz.ch}
}


%





\maketitle
\thispagestyle{fancy}
\pagestyle{fancy}


\begin{abstract}
While distribution networks (DNs) turn from consumers to active and responsive intelligent DNs, the question of how to represent them in large-scale transmission network (TN) studies is still under investigation. The standard approach that uses aggregated models for the inverter-interfaced generation and conventional load models introduces significant errors to the dynamic modeling that can lead to instabilities. This paper presents a new approach based on quantile forecasting to represent the uncertainty originating in DNs at the TN level. First, we aquire a required rich dataset employing Monte Carlo simulations of a DN. Then, we use machine learning (ML) algorithms to not only predict the most probable response but also intervals of potential responses with predefined confidence. These quantile methods represent the variance in DN responses at the TN level. The results indicate excellent performance for most ML techniques. The tuned quantile equivalents predict accurate bands for the current at the TN/DN-interface, and tests with unseen TN conditions indicate robustness. A final assessment that compares the MC trajectories against the predicted intervals highlights the potential of the proposed method.
\end{abstract}

\begin{IEEEkeywords}
Active distribution network, frequency stability, dynamic equivalents, Monte Carlo simulation, quantile forecasts
\end{IEEEkeywords}


%
\IEEEpeerreviewmaketitle

\section{Introduction}
In the last decade, increasing PV generation and smart loads have turned distribution networks (DNs) from passive elements that only consume power into active, intelligent grids that support the transmission network (TN) operation. Literature on different control strategies to enable grid support from decentralized units is widely available \cite{Karbouj}. However, the question of how to aggregate and represent these inverter-interfaced devices in TN level studies is still under investigation \cite{Milano_2018, Calero_2021}.

Recent literature explores various techniques to approach the aggregation of inverter-interfaced units in large-scale TN studies. The authors in \cite{Zali_Milanovic_2013, Conte2017} propose a generic gray box model that consists of a composite load model and inverter-based generation (IBG) in parallel. They employ simple identification techniques to arrive at seventh-order state-space representations. An advanced gray box model is developed in \cite{ChaspierrePhD}, where elaborate models that include detailed representations of protection and support functions are employed for the DN components to formulate dynamic equivalents. While all the literature above focuses on large disturbance and voltage stability, an extension for frequency stability studies is provided in \cite{Dozein_2021}. The authors design a closed-loop identification approach that captures the continuous interaction between the TN and decentralized PV resources. The method is suitable for low inertia systems. However, all of these equivalents are only valid for the operating point they were designed for, and only \cite{ChaspierrePhD} assesses the uncertainty imposed by DNs on the TN level in the derivation of dynamic equivalents \cite{VorwerkPSCC}.  

Black box algorithms trained with machine learning (ML) techniques are another approach for aggregating DNs. The authors in~\cite{Azmy2004} utilize a recurrent neural network (NN) to represent all active resources within a DN and connect it at the TN/DN interface. Batteries are included in \cite{Calero_2021} where two parallel NNs, one for active and one for reactive power, predict the response of distributed batteries at the TN level. However, both studies use small datasets for training and only provide a qualitative assessment instead of quantitative comparison and evaluation of several models. While \cite{Calero_2021} includes frequency and voltage disturbances in the training set, the uncertainty originating from DN parameterization and initial conditions is not reflected in the trained models. 

This work addresses the aforementioned shortcomings and proposes the use of ML-based quantile forecasting for creating robust dynamic equivalents for representation of DN in time-domain simulations and stability studies of low-inertia systems at the TN level. In contrast to point forecasts, which are deterministic and predict the most probable value, quantile forecasts produce predictions with different probabilities and are capable of representing the uncertainty of the data set. A confidence interval for the quantity of interest is hence obtained by combining two different quantile forecasts. While they have been considered in the context of price forecasting, voltage control, and flexibility quantification~\cite{Wan_2014, ThierryPhd, ZUFFEREY2020106562}, the quantile forecasts, to the authors' knowledge, have not been used for dynamic equivalents of DNs. 

First, we generate a large dataset using Monte Carlo (MC) simulations for a test DN subject to a load step at the TN level. The dataset consists of 1000 time-series for different frequency disturbances (i.e., load steps) and DN parameterizations (i.e., load and generation model parameters and operating conditions). Besides employing detailed support and protection functions for IBGs, detailed models of active thermal loads (ATLs) are also included. Second, we comprehensively select, tune, and train point forecast algorithms with the aim of predicting the current at the TN/DN-interface in reaction to a load step. To this end, we consider and compare different ML regressors like linear regression, elastic net regression, gradient boosting trees, and neural networks. Finally, we tune their quantile versions to predict the current at the TN/DN-interface within a given confidence interval. Besides testing the obtained equivalents on the TN parameterization they were trained for, we subject them to other TN conditions and assess their robustness. Finally, the current bands obtained with the tuned ML-based quantile dynamic equivalent are compared to the responses of the MC dynamic simulations.

The remainder of this paper is structured as follows: Section~\ref{sec:process} details the proposed process for tuning and using quantile dynamic equivalents. Then, Section~\ref{sec:gentimeseries} describes the process of generating a rich dataset with MC dynamic simulations on a detailed DN. Section~\ref{sec:pointforecast} details the tuning, testing, and evaluation of the point forecast algorithms, while in Section~\ref{sec:quantileforecast}, the quantile predictors are trained and evaluated. Two additional assessments are performed in Section~\ref{sec:weakTN}. First, the quantile equivalents are tested on a dataset generated with a weak TN equivalent. Then, the current bands obtained through ML-based quantile forecasts are compared to the initial MC simulation results. Section~\ref{conclusion} concludes the paper.

\section{Overview of the Proposed Quantile Dynamic Equivalents}\label{sec:process}
This study aims to develop dynamic equivalents capable of representing DNs in TN stability studies while capturing the uncertainty originating inside the DN. In modern DNs, several sources of uncertainty exist. For instance, model parameters are typically inaccurate, especially for load and renewable generation modeling. In addition, the initial operating conditions of the DN and the load and generation powers are typically uncertain. Finally, some grid codes only specify boundary conditions for protection and control parameters instead of specific setting values~\cite{ChaspierrePhD, IEEE19, Eug}. 

One way of coping with the uncertainty is to use averaged gray box models \cite{ChaspierrePhD}. However, the resulting dynamic equivalents only provide a mean response. In reality the actual system behavior can differ from the mean. In the worst case, the resulting analysis using such models might be optimistic, leading to insecure system operation. To address this issue, the presented work proposes the use of quantile models that build an interval of potential responses in addition to the mean behavior. 

\begin{figure}[t]
    \centering
    \scalebox{0.8}{\includegraphics{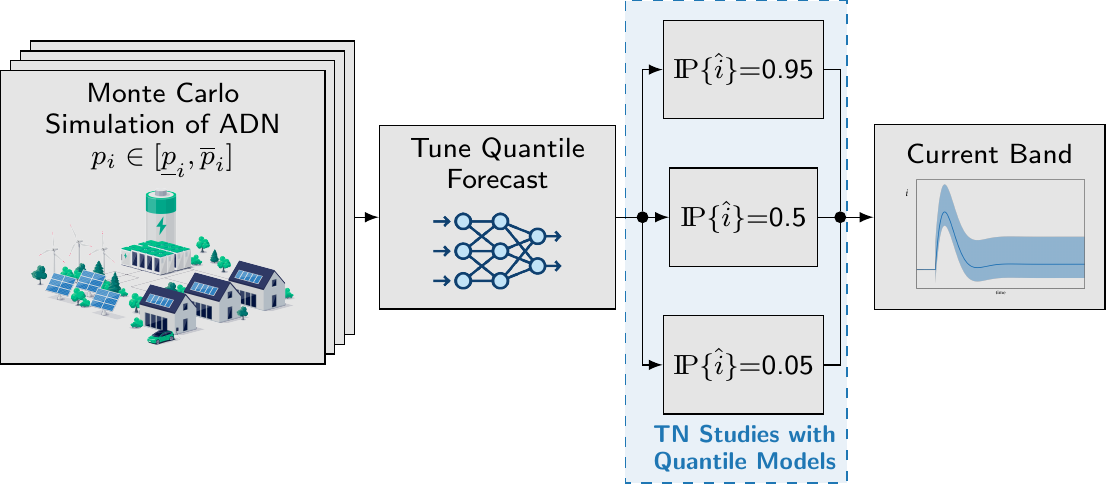}}
    \caption{Overview of the proposed structure for the ADN equivalent in transmission grid stability studies.}
    \label{fig:overviewBB}
\end{figure}

For tuning the underlying ML algorithms, suitable time-series are required. Measured data from actual DNs are one option. Unfortunately, such data is usually biased, including mostly stable cases. However, it is of utmost importance to train ML models on a rich dataset, especially in the case of NNs. Thus, the considered problem requires data for cases with severe, rare disturbances. In this study, MC dynamic simulations on a detailed DN model
are performed to obtain time-series for different frequency events
and a significant number of events that trigger unit protections is included. 


The proposed process is summarized in Fig.~\ref{fig:overviewBB}. First, a MC simulation is used on a DN model with randomly selected parameters. Based on this generated conditions, time-series of frequency and voltage response are generated based on dynamic simulations for different events that excite a dynamic response. In the next stage, the generated time-series data are used to train quantile forecasting models based on voltage and frequency data. The trained quantile dynamic equivalent produces three separate time-series for the DN current $i$ at the point of common coupling (PCC) per disturbance. While one of them predicts the most probable ($\mathbb{P}=0.5$) behavior, the remaining two estimate a lower and upper bound with selected confidence, e.g. 90~\% as shown in Fig.~\ref{fig:overviewBB}. These quantile current bands could be employed in TN stability studies and result in frequency and/or voltage bands that represent the DN uncertainty at the TN level.

\section{Generating Time-Series}\label{sec:gentimeseries}
In this section, the modeling of the DN components and the uncertainty of the DN model is described. Then, the methodology for obtaining the time-series used for the training phase is showcased.

\subsection{Grid Model}
The employed grid model is based on \cite{VorwerkPSCC}, which has been extended for this work. Standard models are used for the lines, transformers, induction motors, and synchronous machines. For the IBGs and ATLs, recently developed models that comply with modern grid support requirements are used~\cite{IEEE19, Eug, AusCode2}. 


\subsubsection{Lines}
Under the phasor approximation, the grid is modeled with the following system of algebraic equations $\underline{\boldsymbol{i}} = \boldsymbol{Y} \underline{\boldsymbol{v}}$, where $\underline{\boldsymbol{i}}$ represents the vector of complex current injections at the system nodes and $\underline{\boldsymbol{v}}$ is the vector of nodal voltage phasors. The admittance matrix $\boldsymbol{Y}$ includes the line and transformer impedances at nominal frequency $f_n$. Note that all lines and potential transformers are modeled with standard $\Pi$-equivalents as described in \cite{Kundur1994}.

\subsubsection{Background Load}
Each DN node might connect different types of load and generation. While ATLs and IBGs are modeled separately, all other load and generation is lumped into a background load model. This generalized background load consists of a static and a dynamic component. While the former part is modeled by an exponential load model \cite{Kundur1994}, a single case induction machine (IM) with third-order formulation represents the latter \cite{Kundur1994}. 

Unlike \cite{VorwerkPSCC}, where the ATL power and initial load per node $P_0$ are assumed to be known, in this study, only the latter is known while the load distribution among thermal, static, and dynamic load is uncertain. Thus, the initial static background load power $P_{l,0}$ is formulated as:
\begin{align}
    P_{l,0} = (1-f_\mathrm{im}-f_\mathrm{atl})\cdot  P_0,
\end{align}
where $f_\mathrm{im}$ and $f_\mathrm{atl}$ are the initial load shares of the IM and ATL, respectively.

\begin{figure}[!t]
\begin{minipage}{\columnwidth}
    \centering
    \scalebox{0.85}{\includegraphics{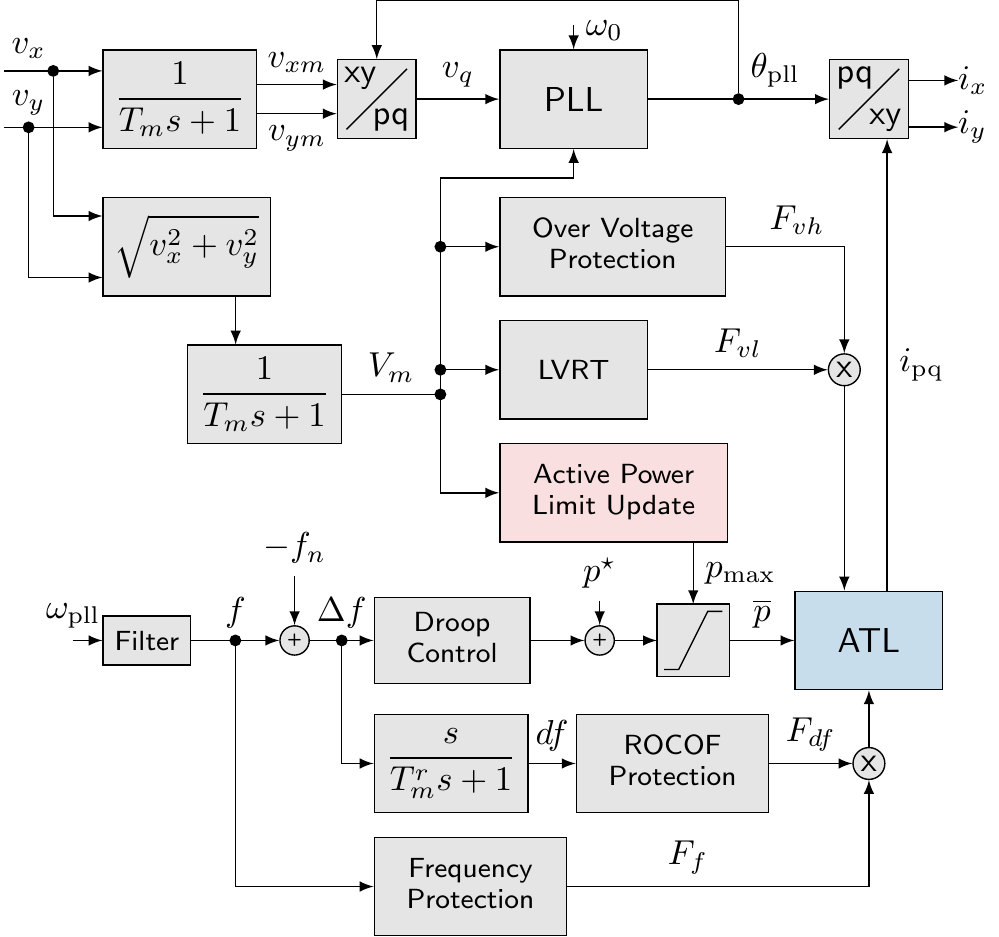}}
    \caption{Overview of the thermal load model. Blocks that deviate from the IBG model are highlighted.}
    \label{ATL_overview}
    \vspace{10pt}
    \scalebox{0.85}{\includegraphics{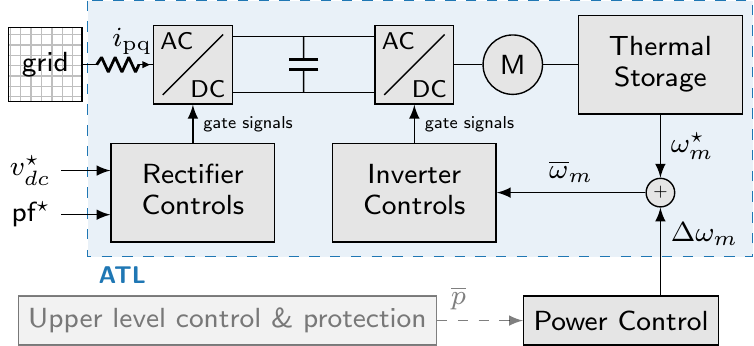}}
    \caption{Overview of the detailed inner thermal load model.}
    \label{ATL_detail}
\end{minipage}
\end{figure}

\subsubsection{Active Thermal Loads}
The ATL model represents various compressor-based thermal loads, like heat pumps, refrigeration, and air conditioning devices. While it ignores the switching of the internal power electronics, it captures the primary behavior of the devices. Figure~\ref{ATL_overview} provides an overview of the outer control and protection layer. Besides incorporating the latest grid support and protection functions required by modern grid codes,
the inherent dynamics of the thermal load are considered in detail. 

The inner ATL model, marked by the blue block in Fig.~\ref{ATL_overview} and detailed in Fig.~\ref{ATL_detail}, captures the inverter, motor, and thermal load dynamics as well as their various controls. While fast controls like the rectifier and inverter controls are implemented, the temperature control loop is omitted. The latter time scales range in hours, while those of interest for this study range in seconds. Refer to \cite{VorwerkPSCC} for a detailed model description.

Unlike the implementation in \cite{VorwerkPSCC} where most ATLs share the same nominal power $S_b$, in this implementation, it is adjusted by the load factor $\mathrm{LF}$ to consider different initial loading conditions of the ATLs during MC simulations. The base power $S_b$ of the ATL is adjusted according to $S_b = f_\mathrm{atl} P_0 / \mathrm{LF}$, where the load factor is randomly chosen.

\begin{figure}
    \centering
     \scalebox{0.85}{\includegraphics{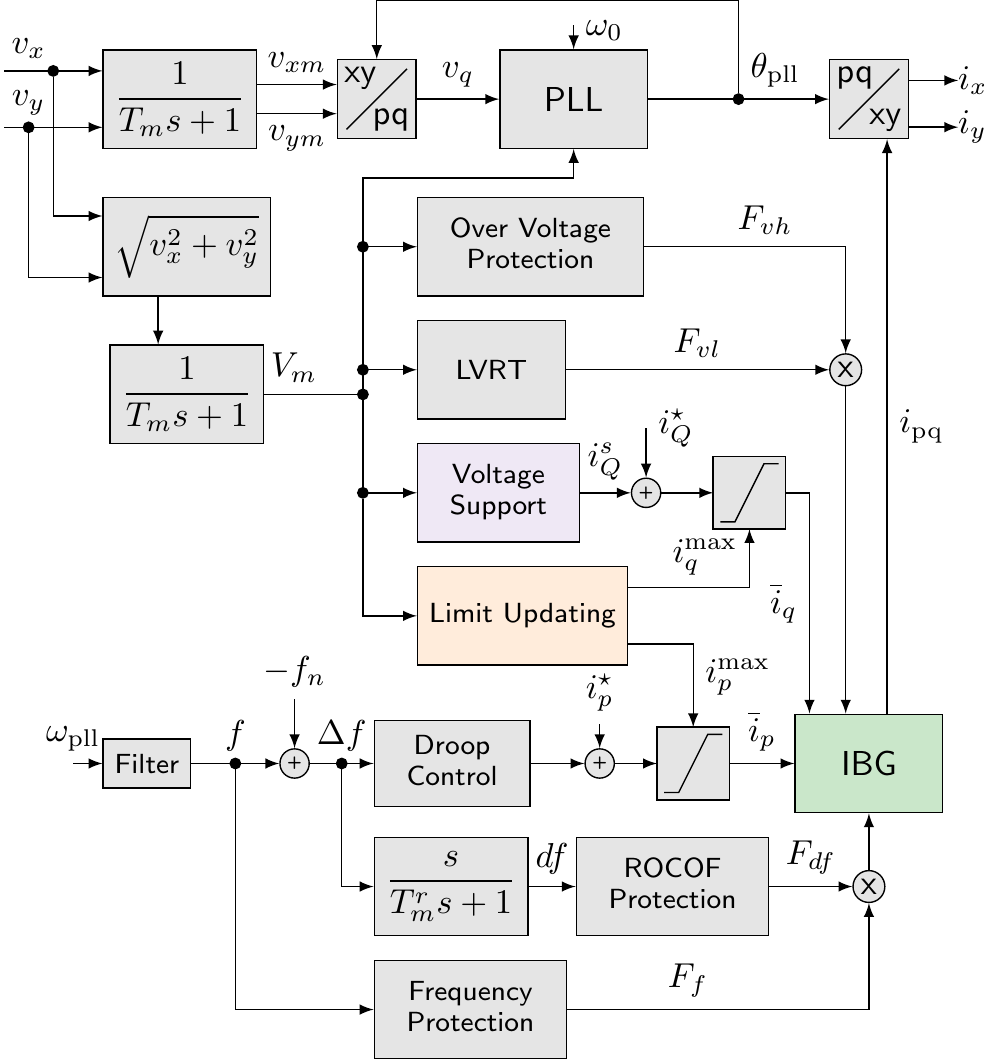}}
    \caption{Overview of the IBG model. Blocks that are different from the ATL implementation are highlighted.}
    \label{fig:IBG}
\end{figure}

\subsubsection{Distributed Generation}
Like the ATL model, the IBG model complies with modern grid codes and implements complementary protection and grid supporting functions. Figure~\ref{fig:IBG} provides an overview of the grid-level controls and protections. Despite the similarities, there are some particular differences: while the ATL model only manipulates its active power consumption to aid voltage and frequency, the IBG model also supports the grid through reactive power control. To this extent, reactive power control priority and a block that updates the limits for reactive and active power are included. 

The inner IBG model, marked in green in Fig.~\ref{fig:IBG}, only considers the outer loop dynamics of its controllers. Therefore, it assumes good tracking performance of the current control and neglects the fast internal dynamics. The inverter current is assumed to follow a first-order transfer function \cite{ChaspierrePhD}. The active and reactive current references are computed from the active and reactive power setpoints. They are altered by the various outer control and protection functions. The implemented model complies with \cite{AusCode2, Eug, IEEE19} and is described in detail in \cite{VorwerkPSCC}.

\subsubsection{Transmission Grid Equivalent}
To capture frequency dynamics, an equivalent synchronous machine (SM) model is used as the TN equivalent. The SM is incorporated with a fifth-order model \cite{Kundur1994} and connected to the TN/DN-interface via an additional line for adjusting the short-circuit power and $R/X$-ratio of the respective TN. In addition, the governor is represented by the standard IEEE~TGOV1 model from \cite{IEEE_gov}, with $T_2=T_3=0$, and the minimum and maximum voltages are set to $V_\mathrm{min}=0~\text{p.u.}$ and $V_\mathrm{max} = 1 \text{ p.u.}$, respectively. The standard IEEE~AC1A model from \cite{IEEE_exc} is employed for the exciter.


\subsection{Capturing Uncertainty}


One of the key aims of this study is to use quantile dynamic equivalents to represent the uncertainty caused by ADNs in TN stability studies. Thus, several of the ATL model parameters (i.e. motor inertia, friction constant, anchor resistance, outer control loop parameters)  are included in the uncertainty set. The inner control loops are significantly faster and thus are not included in the uncertainty set. Similarly, the outer control loop and protection parameters of the IBGs are assumed uncertain as proposed in \cite{ChaspierrePhD, ChaspierrePSCC18, VorwerkPSCC}.

In addition to parametric uncertainty, the initial operating point also introduces uncertainty. This study considers one operation point, assuming the initial nodal load is known. However, the initial load distribution is varied, i.e., the share of static, dynamic, and thermal load consumption. In addition, the load factor of the ATLs is also included in the set of uncertain parameters.


Table~\ref{tab:Par} provides an overview of all parameters of the uncertainty set and their respective ranges. The initial thermal load share relates to realistic ranges for German load data \cite{VorwerkPSCC, ThermalData}. Since the statistical distribution of the uncertain parameters is unknown, a uniform distribution is assumed during MC simulations as suggested in \cite{Mooney}. 

\begin{table}[t]
  \centering
  \fontsize{7}{7} 
  \caption{Variation of Parameters.} \label{tab:Par}
  \scalebox{0.75}{
  \begin{tabular}{p{2.5cm} l c p{3cm} l}
    \toprule
  \multicolumn{2}{l}{\textbf{Active Thermal Load}} & & \multicolumn{2}{l}{\textbf{Induction Machines}} \\ \cmidrule{1-2} \cmidrule{4-5}
  PLL delay $\tau_\mathrm{pll}$ & $[0.05,0.1]$ s & & Stator resistance $r_s$ & $[0.03,0.13]$ p.u.\\
  Power control & \multirow{2}{*}{$[0.01,0.03]$ s} & & Rotor resistance $r_r$ & $[0.03,0.13$ p.u. \\
  time constant $\tau_p$ & & & Magn. inductance $l_m$ & $[2.5,4]$ p.u. \\
  Inertia constant $H$ & $[0.03,0.5]$ s & & Stator inductance $l_s$ & $[0.07,0.15]$ p.u. \\
  Friction constant $b$ & $[0.0005,0.002]$ p.u. & & Rotor inductance $l_r$ & $[0.06,0.15]$ p.u. \\
  Input resistance $r_t$ & $[0.005.0.05]$ p.u. & & Inertia constant $H$ & $[0.2,1]$ s \\
  Input impedance $l_t$& $[0.1,0.9]$ p.u.& & Load factor LF & $[0.4,0.6]$ \\ 
  Anchor resistance $r_a$ & $[0.01,0.1]$ p.u. & & Power factor $\cos \varphi$ & $[0.85,0.95]$ \\ 
  Initial share $f_\mathrm{atl}$ & $[0.01, 0.4]$  & & Initial share $f_\mathrm{im}$ & $[0,0.2]$ \\ 
  Load factor $LF$ & $[0.3,1.3]$ p.u. & & & \\ \cmidrule{1-2} \cmidrule{4-5}
  \multicolumn{2}{l}{\textbf{IBG Units}} & & \multicolumn{2}{l}{\textbf{Static Loads}} \\ \cmidrule{1-2} \cmidrule{4-5}
   PLL delay $\tau_{pll}$ & $[0.05,0.1]$ s& & Load exponent $\alpha$ & $ [1,2]$ \\
   Current control  & \multirow{2}{*}{$[0.01,0.03]$ s} & & Load exponent $\beta$ & $[1.5,3]$ \\ 
   time constant $\tau_i$ & & & &  \\
   Ramp limit $\frac{di_p}{dt}$ & $ [0.2,0.5]\frac{\text{p.u.}}{s}$ & & \\\bottomrule
  \end{tabular}}
  \vspace{-5pt}
\end{table}
\begin{figure}
    \centering
    \scalebox{0.72}{\includegraphics{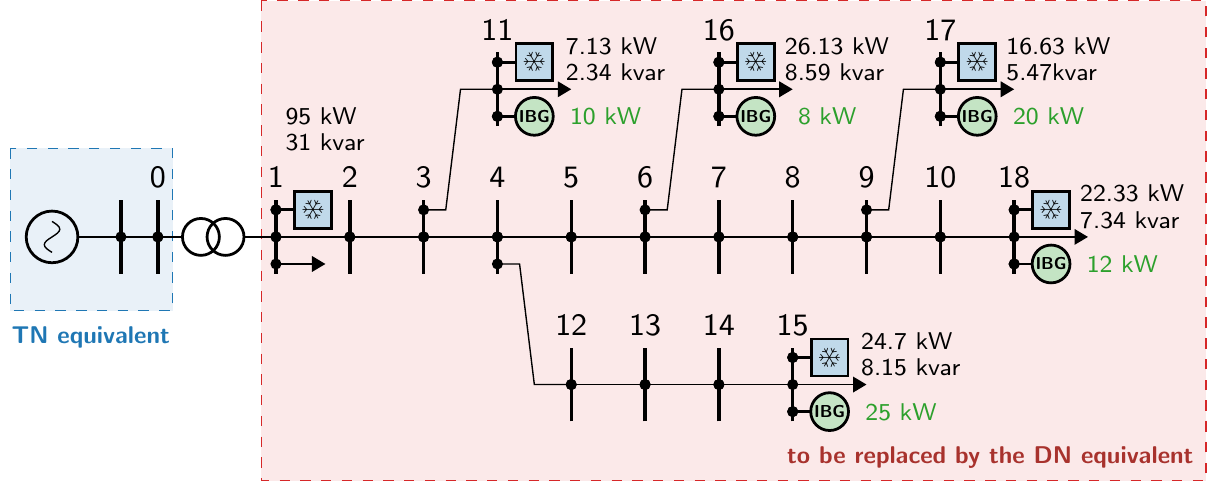}}
    \caption{Overview of the applied and adjusted CIGRE 18 Bus test case. Initial generation per node is given in green, while the initial nodal consumption including static, thermal and dynamic loads is provided in black.}
    \label{fig:TestSystem}
\end{figure}

\subsection{Obtaining the Time-Series}
The required time-series are obtained by MC dynamic simulations on the CIGRE European 18 bus residential low voltage network published in \cite{TestCase}. First, the modifications of the test system and the considered test cases are discussed. Then, the implementation and obtained time-series are presented.

\subsubsection{Test System}\label{sec:testsystem}
All line, transformer, and grid equivalent parameters are taken from \cite{TestCase} and the loads are reduced to $50\%$ of the initial loading, keeping the nodal voltages within $\Delta V=\pm0.1$~p.u. bounds. The initial load and generation per node are stated next to the corresponding node in Fig.~\ref{fig:TestSystem}. While the load distribution varies per MC simulation, the PV generation remains constant. Note that all ATL and IBG units operate at unity power factor. The protection and support settings comply with \cite{AusCode2} and are listed in \cite{VorwerkPSCC}. 
\begin{figure}[t!]
    \centering
    \includegraphics{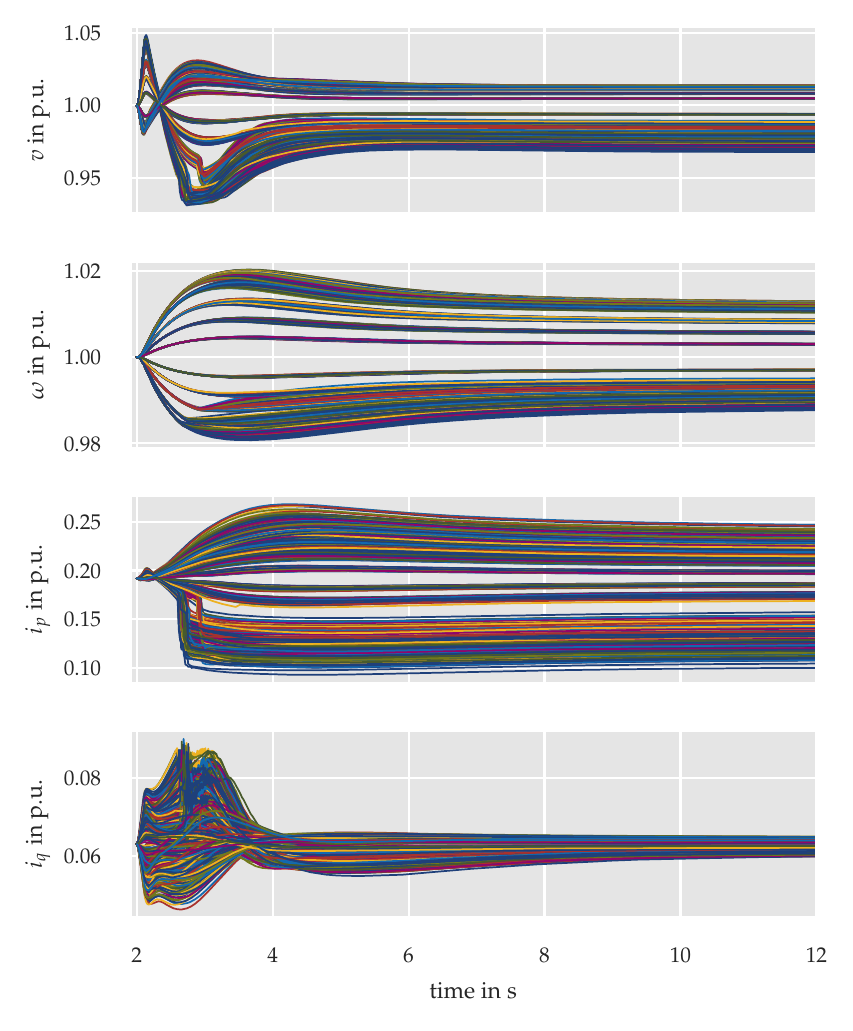}
    \caption{Generated data: 1000 time domain simulations with ten different load step changes and 100 sets of random parameters for the strong grid case.}
    \label{fig:Data}
\end{figure}

The equivalent TN corresponds a strong grid with a short-circuit power of $150~\text{MVA}$ and \SI{6}{\second} of inertia. The equivalent SM nominal power equals that of the transformer connecting the DN. Thereby, uncertainty in the DN is reflected in TN quantities like the system's frequency. It is similar to assume that the TN connects several DNs in parallel that all respond to a TN disturbance in the same fashion. A constant power load is connected to the TN bus for simulating load step changes. 

For each MC simulation, the parameters are drawn from the ranges in Table~\ref{tab:Par}. One MC simulation set contains 10 different frequency events. In other words, each parameter set is subjected to ten load steps ranging within $\Delta P_l=\pm$225~kW. Overall, 100 parameterizations are drawn, resulting to 1000 time-series for ML training and testing. The load step always occurs at $t=\SI{2}{\second}$ and simulations are executed for \SI{12}{\second}.

\subsubsection{Implementation}
The test system and corresponding models are implemented in the dynamic simulation software PyRAMSES~\cite{2016JAristidou}. All models have been validated in~\cite{VorwerkPSCC, ChaspierrePhD}. The sampling time for a time-series varies since a variable step-size solver is employed for the simulations. Thus, each of the 1000 time-series contains between 500 and 2000 points. 

\begin{figure}[t!]
    \centering
    \includegraphics{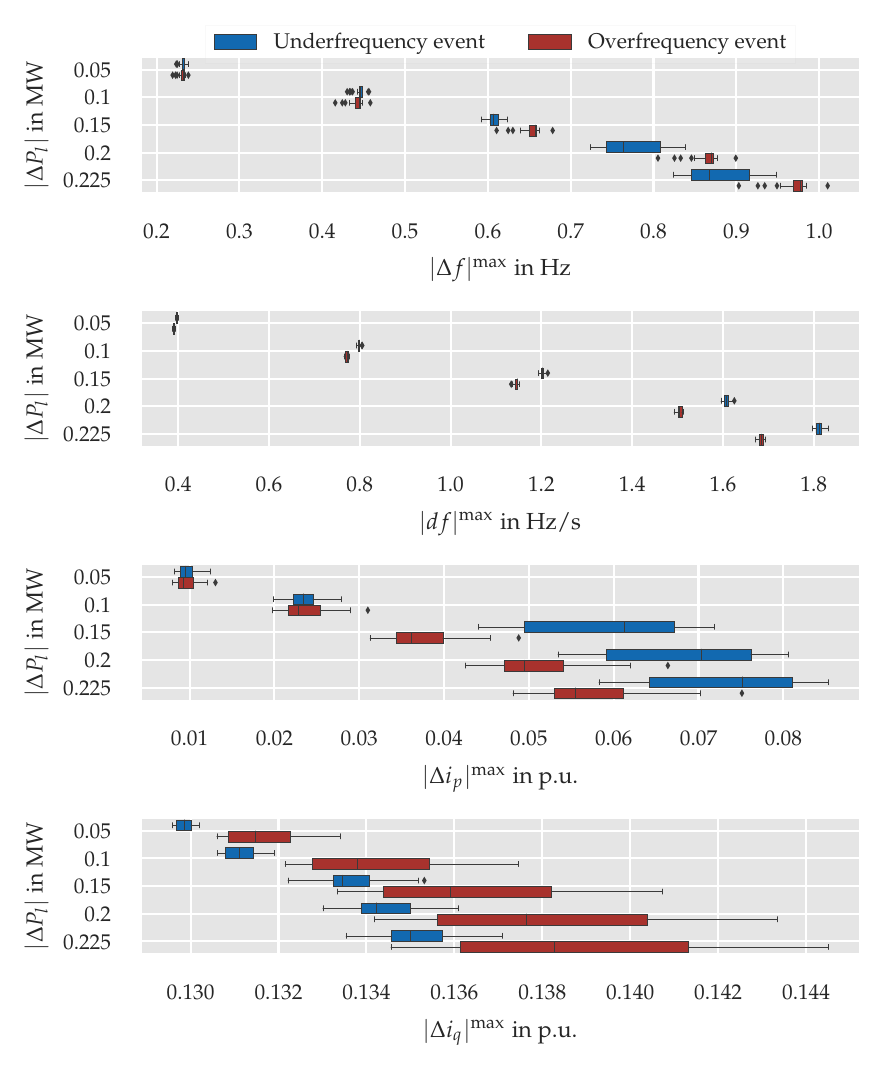}
    \caption{Boxplot of the absolute maximum (from top to bottom) frequency deviation, ROCOF, active and reactive current deviations for the ten different load step changes and 100 random parameter sets.}
    \label{fig:boxplot}
\end{figure}

\subsubsection{Overview of Obtained Time-Series}
An overview of the 1000 obtained time-domain simulations is provided in Fig.~\ref{fig:Data}, where $v$ is the voltage magnitude at bus 1 (PCC), $\omega$ is the TN equivalent's frequency, and $i_p$ and $i_q$ represent the active and reactive current across the TN/DN-transformer. Note that the current is positive if flowing from the TN into the DN. 

The variance present in the time-series is indicated in Fig.~\ref{fig:boxplot}. It contains boxplots of the maximum absolute frequency deviation $\abs{\Delta f}^{max}$, the maximum absolute ROCOF $\abs{df}^\mathrm{max}$, as well as the absolute maximum active and reactive current deviations, $\abs{\Delta i_p}^{\mathrm{max}}$ and $\abs{\Delta i_q}^\mathrm{max}$, for each load step $\abs{\Delta P_l}$. Except for the ROCOF, which generally shows small deviations from its average, the variance in frequency and current deviations increases significantly for larger load steps. The asymmetric behavior for under- and overfrequency events mainly stems from the asymmetric protection settings required by the considered grid code~\cite{AusCode2}. 

\begin{figure}
    \centering
    \scalebox{0.8}{\includegraphics{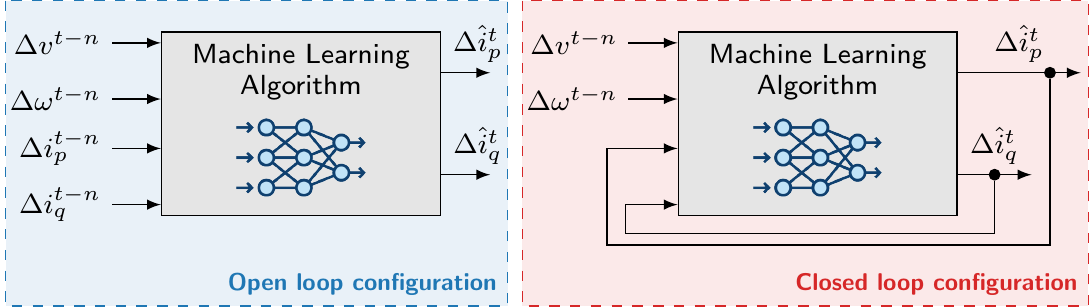}}
    \caption{Different configurations: open loop (left) is used during training and closed loop (right) during validation and testing. Note that $n>0$ must hold for the current features.}
    \label{fig:configuration}
\end{figure}

\section{Point Forecasts}\label{sec:pointforecast}
This sections discusses several point forecast ML algorithms and introduces multiple quantification scores to compare the algorithms. Moreover, the feature selection and hyper-parameter tuning techniques for each algorithm are presented.

\subsection{Methods}
Different ML methods are implemented and tuned to identify the best algorithms for the problem under consideration: Linear regression (Linreg), regularized linear regression (Elnet), support vector regression (SVR), gradient boosting trees (GBTs), and neural networks (NNs). Since preliminary testing has revealed extensive training times and poor performance for the SVR and since the authors are not aware of a quantile counterpart, it is not discussed further.

The selected target variables for each algorithm are the active and reactive currents (${i}_p$ and ${i}_q$) over the TN/DN-transformer. The currents are chosen instead of power to ensure a separation of features and targets. The currents could be formulated in the stationary grid reference frame, the ($xy$)-frame. But, in such a scenario the resulting DN equivalent is prone to changing voltage angles at the TN level. 

The potential input features for each method include the voltage magnitude at the TN/DN-interface $v^{t-n}$ and the TN frequency $\omega^{t-n}$. The superscript ${t-n}$ indicates a $n$-times shifted version of the input, i.e. historical values of the signal. For each algorithm, shifts of $n\in[0,10]$ are considered. Note that $n=0$ indicates that the input feature is measured at the same point in time as the target is predicted. Although the power step change at TN level is available during power system stability studies, it is omitted to enhance the performance when placing the DN equivalent in a different TN. 

Instead of using the actual magnitude of each quantity, deviations from steady-state, indicated by $\Delta$-notation, are computed to increase the robustness against changing initial conditions at TN level. In addition, the data are scaled by standardization, i.e. centering around zero and scaling to unit variance, for training, validation and testing. Nonetheless, all data are transformed back before performance quantification. 

Preliminary studies indicate a drastic increase in the performance of all ML techniques when adding former predictions as input features for the new prediction. Thus, $n$ former predictions  $\hat{i}_p^{t-n}$ and $\hat{i}_q^{t-n}$ are also considered as features. Note that for these it holds: $n\in[1,10]$.

Since former current values are used as input features, training, validating and testing of the algorithms are performed using different configurations. During training, current predictions are not available such that actual time-shifted currents are given as input features. This is referred to as open-loop configuration. During validation and testing, the former predictions of the currents are available and fed back to the algorithm as input features for the next predictions. This is referred to as closed-loop configuration. Both configurations are illustrated in Fig.~\ref{fig:configuration}.

The chosen algorithms assume that the targets do not affect the input features. But in real power systems, that is not the case, i.e. the current drawn at the TN/DN interface affects the voltage and frequency at the TN level. While this effect is limited for strong grids, it is potentially significant for weak TNs and future work should focus on integrating the link between targets and input features in the modeling process.

\afterpage{
\begin{table}[t!]
    \centering
    \caption{Chosen features and hyper-parameters for each model.}
    \renewcommand{\arraystretch}{1.1}
    \footnotesize
    \label{tab:MLpars}
    \begin{tabular}{l p{0.55cm} p{5.5cm}}
    \toprule
     \textbf{Algorithm}  &   \textbf{n}$^\text{\textbf{a}}$& \textbf{Hyper-parameters}\\ \midrule
    linreg     & $[0,1]$ & \\ 
    elnet & $[0,10]$ & $\alpha = 0.1 $,  $\rho_{i_p} = 0.12$, $\rho_{i_q} = 0.24$ \\ 
    gbt $i_p$ & $[0,1]$ & 300 estimators, min samples leaf: 11, \newline max depth: 5, max features : auto\\ 
    gbt $i_q$ &  $[0,1]$  & 225 estimators, min samples leaf: 6, \newline max depth: 5, max features : auto\\ 
    NNt & $[0,5]$ & adam, dropout:~0.01, batch size:~300, 40~epochs \\
    NNb & $[0,5]$ & adam,  dropout:~0.01, batch size:~200, 40~epochs\\\bottomrule
    \end{tabular}
    \begin{tablenotes}
    \scriptsize
            \item[a] \textbf{a} For all algorithms the features are the voltages $\Delta v^{t-n}$, frequencies $\Delta \omega^{t-n}$ and former predictions $\Delta \hat{i}_p^{t-n}$, and $\Delta \hat{i}_q^{t-n}$. Note that for the two former predictions the following must hold $n>1$.
    \end{tablenotes}
\end{table}
}

In the following, the fine-tuning of each algorithm is presented. For this purpose, 120 time-series are used for training and 80 series are included in the validation set. All subsets contain data for all disturbances. The ML techniques are implemented in Python using scikit-learn \cite{scikit-learn} and keras~\cite{keras}. Feature selection is performed with the default model parameters in the first stage, while grid searches determine the best hyper-parameters in the second stage. Generally, the procedures in \cite{CourseraML, sklearn_tutorial, DataCamp_sk, DataCamp_keras} are followed. Note that if required, performance on $i_p$ is prioritized because the obtained models are meant to be applied in frequency stability studies. Table~\ref{tab:MLpars} lists the chosen features and hyper-parameters per algorithm. 

\subsubsection{Linear Regression}
Linreg is used as the benchmark model and is the simplest of the implemented algorithms. The target variable $\hat{y}$ is computed by a linear combination of the features $x$ with:
\begin{align}
\hat{y} = \sum_{i=1}^{m} \beta_i x_i,
\end{align}
where $m$ is the number of features and $\beta$ are the model coefficients that are obtained by minimizing the following least-squares objective: 
\begin{align}
    J_\mathrm{lin} = \min_\beta || y-X \beta||^2,
\end{align}
where $y$ is the true value. 

Linreg does not have any hyper-parameters and only the features need to be selected. If more than one former feature ($n>1$) is used, the closed loop performance on the validation set is poor. Thus, six features are chosen: $\Delta v^t$, $\Delta v^{t-1}$, $\Delta \omega^t$, $\Delta \omega^{t-1}$, $\Delta i_p^{t-1}$ and $\Delta i_q^{t-1}$, where $\Delta$ indicates that deviation from steady-state are used. The algorithm is implemented in scikit-learn using the LinearRegression function \cite{sklearn_tutorial}.

\subsubsection{Elastic-Net Regression}
Elnet regression applies the same linear combinations as Linreg. Nonetheless, two regularization terms are added to the cost function of the minimization problem. It combines Lasso and Ridge regression (i.e., L1 and L2 regularization) and is capable of handling correlated features. The cost function is:
\begin{align}
    J_\mathrm{el} =  \min_\beta \frac{1}{2n_s} \norm{X \beta-\beta}_2^2+\alpha \rho \norm{\beta}_1 + \frac{\alpha(1-\rho)}{2}\norm{\beta}_2^2
\end{align}
where $\alpha$ controls the effect and $\rho$ the balance between both regularization types. $n_s$ is the number of samples. The algorithm is implemented in scikit-learn using the ElasticNet function \cite{sklearn_tutorial}.

After selecting the features, a grid search is performed to select the best hyper-parameters. Both, $\alpha$ and $\rho$ are considered at this stage, and the ranges are narrowed through several iterations. In general, best scores are achieved for $\alpha\leq0.1$, indicating that linreg describes the data better than regularized regression. 

\subsubsection{Gradient Boosting Tree}
GBT regression is an ensemble method. As such, it combines several weak learners to increase the robustness compared to a single estimator. In boosting, the base predictors are built sequentially and the bias of the combined estimator is reduced. 

Similar to other boosting methods, GBT is built in a greedy fashion, and each newly added tree is fitted to minimize a sum of losses given by the previous ensemble. The loss function can be arbitrarily chosen, but for the presented work the mean squared error is selected. The reader is referred to \cite{sklearn_tutorial} for details and the mathematical formulation. The implemented algorithm is the GradientBoostingRegressor method.

While all potential features are considered during feature selection, only some parameters of the algorithm are considered during hyper-parameter tuning \cite{Jain_2016}. The number of estimators controls the amount of boosting stages to perform. It forms a trade-off with the learning rate that shrinks the contribution of each tree. In this work, the number of estimators is adapted, while the learning rate is kept at 0.1. Regarding the tree parameters, the maximum depth of each tree, the minimum number of observations per leaf node and the number of features to consider when looking for the best split are included in a grid search. The best model parameters differ for the two targets. 

\afterpage{
\begin{table}[h]
    \centering
    \footnotesize
    \renewcommand{\arraystretch}{1.2}
    \caption{Training duration for the different algorithms. }
    \label{tab:Durations}
    \begin{tabular}{l| c c c c c }
    \hline
    \textbf{Algorithm}     & linreg & elnet  & gbt & NNt & NNb   \\ \hline
    \textbf{Duration in s}  & 0.297 & 128.9  & 937.7& 1080.5  &  1571.3  \\  \hline
    \end{tabular}
\end{table}
\vspace{-5pt}
\begin{figure}[h]
    \centering
    \scalebox{0.88}{\includegraphics{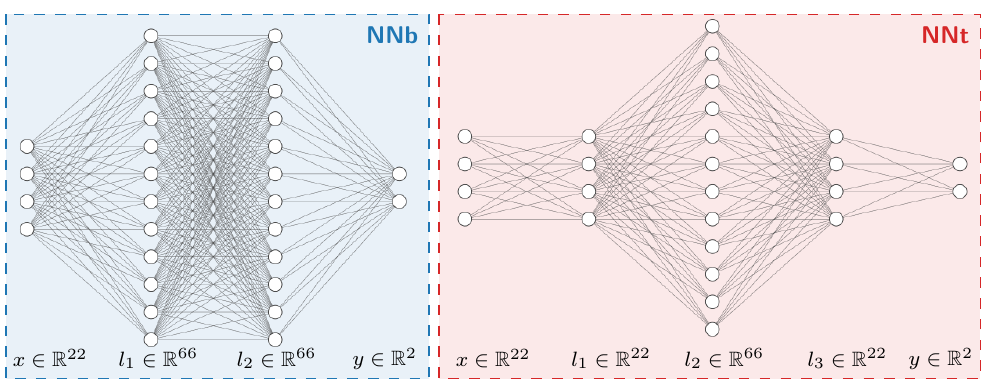}}
    \caption{Architectures of the neural networks with best performance during hyper-parameter tuning. The dimensions of the input $x$, the $i$-th hidden layer $l$ and outputs $y$ are stated below each layer.}
    \label{fig:NN_architectures}
\end{figure}
}

\subsubsection{Neural Networks}
Neural networks are well known ML algorithms inspired by the architecture of human brains. They can approximate any non-linear function via a weighted linear combination of nodes (neurons) that are organized in layers \cite{CourseraML, keras}. In this study, block and trapezoidal architectures, referred to as NNb and NNt, are examined. Note that a sequence of batch normalization and dropout layers follows every hidden layer. While the hidden layers are implemented using Dense layers in keras, the dropout and normalization are implemented with the BatchNormalization and Dropout layers. Preliminary studies demonstrate better performance when one NN predicts both currents simultaneously compared to two NNs predicting one target each. During hyper-parameter tuning, relu, linear, sigmoid and tanh activation functions are analyzed. At the same time, the dropout percentage, batch size and training epochs are optimized using a grid search.

Considering that two architectures result in similarly good validation scores, both of them are included in the upcoming testing stage. The resulting architectures are displayed in Fig.~\ref{fig:NN_architectures}, indicating the relations of features to nodes. In the figure, each neuron of the input layer represents the $t-n$~features of one quantity, e.g. $\Delta v^t$ to $\Delta v^{t-n}$. Furthermore, the first hidden layer of the NNb architecture consists of three-fold the number of features, while the first hidden layer of the NNt has the same number of neurons as it has inputs. 

\subsection{Evaluation Metrics}
This section presents the evaluation metrics. The R2-score, the root mean square error~(RMSE) and the mean absolute percentage error~(MAPE) are included. While these scores are commonly applied for regression problems, they do not assess if the algorithm behaves as a persistence model and basically produces a time shifted version of the former prediction to compute the next target \cite{Flovik_2018}. Thus, to evaluate the predictive power of the models, time-differenced data is analyzed.

\subsubsection{R2-Score}
The R2-score, or coefficient of determination~$R^2$, is a key evaluation metrics for regression problems. It represents the proportion of variance in the targets that is explainable by the independent variables in the model. Thereby, it provides a measure of how well unseen data are likely to be predicted by the model. While the best score is 1, no lower bound exists. When it is zero, the model outputs the mean value of the test set. The score is defined as:
\begin{align}
    R^2(y,\hat{y}) &= 1-\frac{ \sum_{i=1}^{n_s} (y_i-\hat{y}_i)^2}{\sum_{i=1}^{n_s} (y_i-\overline{y}_i)^2}, \quad \overline{y} = \frac{1}{n_s}\sum_{i=1}^{n_s} y_i
\end{align}
where $n_s$ is the number of samples. $\hat{y}_i$ and $y_i$ are the predicted and real value of $i$-th data point.

\subsubsection{Root Mean Square Error (RMSE)}  
The RMSE is a frequently used measure that penalizes large errors:
\begin{align}
    \mathrm{RMSE}(y,\hat{y}) =  \sqrt{\smash{\frac{1}{n_s}} \sum_{i=1}^{n_s} \left(y_i-\hat{y}_i\right)^2}.
\end{align}
The lower the RMSE the better the performance, while zero indicates a perfect fit. 

\subsubsection{Mean Absolute Percentage Error} The MAPE accounts for the absolute difference between predicted and real data:
\begin{align}
    \mathrm{MAPE}(y,\hat{y}) =  \frac{1}{n_s} \sum_{i=1}^{n_s} \left \lvert \frac{y_i-\hat{y}_i}{y_i} \right\rvert.
\end{align}
A low MAPE indicates better performance, while the lower bound is zero. Because the MAPE normalizes the error, it is capable of comparing predictions of different magnitudes.

\subsubsection{Time-difference Plots and Score} 
Following the suggestions in \cite{Flovik_2018}, time-difference plots are generated to assess how well the models predict the difference in the target between one time step and the next, rather than the data directly. For each data-point the change in the target is computed by subtracting the target of the former time step from the current one with $\Delta y^t =y^{t}-y^{t-1}$. The same is repeated for the predictions $\Delta \hat{y}^t = \hat{y}^{t}-\hat{y}^{t-1}$. Representing them in scatter plots illustrates the predictive power of the models. Ideally, the observation and prediction differences are equal for each time step, meaning that all points ($\Delta y^t, \Delta \hat{y}^t$) lie on a straight line.  

\subsection{Comparing the Algorithms}
For the final comparison, all models are trained on 800 training time-series, including the former training and validation sets used for fine-tuning the algorithms. The remaining 200 time-series are used for testing the performance in closed-loop fashion. Table~\ref{tab:Durations} provides an overview of the training times for all the time-series in the training set. Note that the training is performed on a 64-bit Windows server with an Intel Xeon Gold 6154 CPU at 3 GHz and one CPU was used.  While linear and elnet regression are trained quickly, GBT and NNt take significantly longer, while NNb is the slowest algorithm.

\begin{figure}[!t]
\centering
\subfloat{\includegraphics[width=\linewidth]{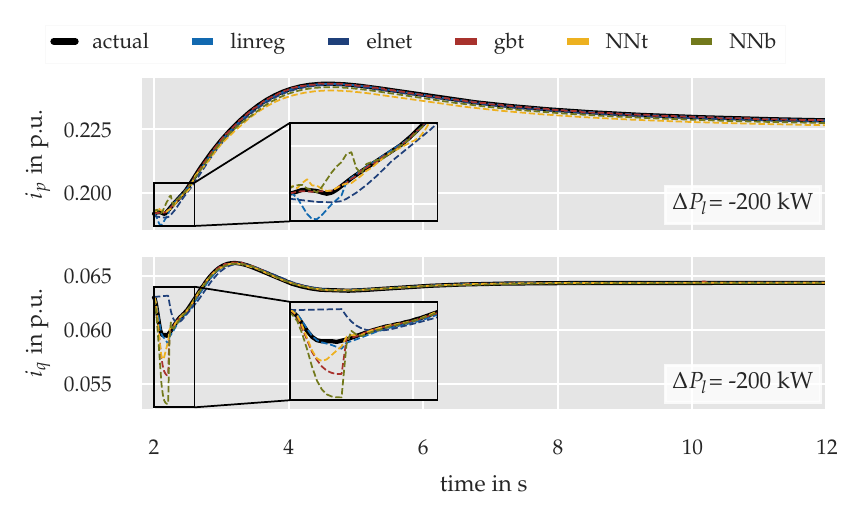}}%
\vspace{-5pt}
\subfloat{\includegraphics[width=\linewidth]{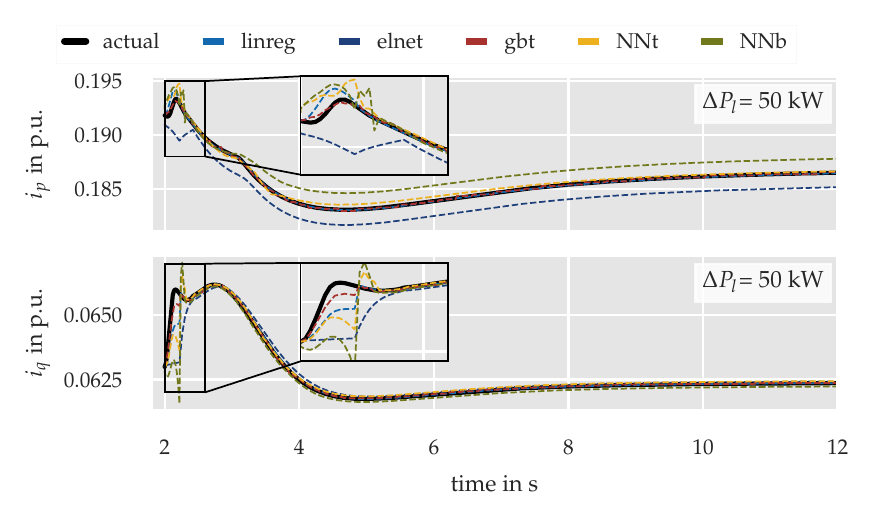}}
\vspace{-5pt}
\subfloat{\includegraphics[width=\linewidth]{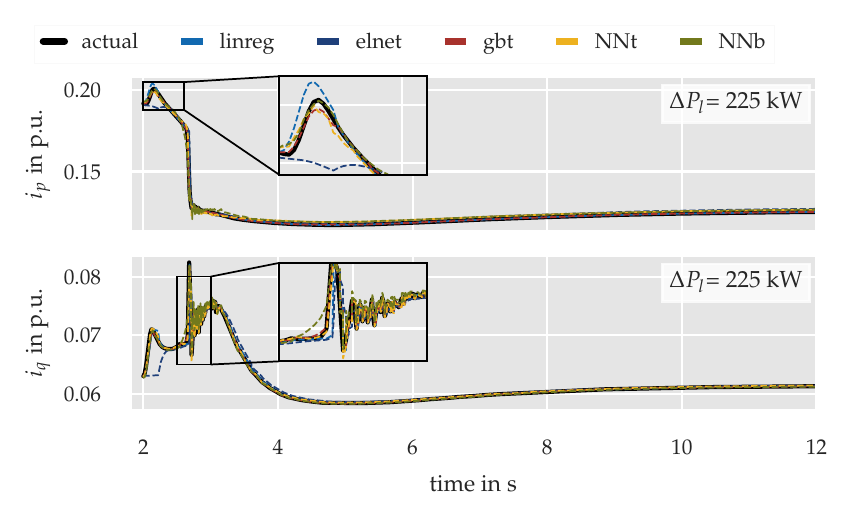}}
\caption{Time domain performance of all algorithms for three different time-series of the test set. The corresponding power step $\Delta P_l$ used to cause the frequency event is provided in each plot. Note the different scaling of the ordinates.}
\label{fig:TD_plots}
\vspace{-10pt}
\end{figure}
Fig.~\ref{fig:TD_plots} shows exemplary time-domain performances for some series of the test set. While the overall shape of the response is generally retrieved for both currents, mismatches occur during the transients. Additionally, deviations seem to occur in the steady-state behavior of the elnet and NNb model for the second case. However, the load step and hence the control reaction, in this case, is small. Mismatches of similar magnitude are observed for other test cases with higher control reactions but invisible in the graphs. For the provided samples, GBT and NNt seem to provide the best prediction for $i_p$ and only exhibit significant mismatches for $i_q$ transients. 

Fig.~\ref{fig:scores} summarizes the closed-loop performance on the entire training and test sets. Note that the performance on the test set is mainly relevant since it shows how the different models perform on unseen data. While for the train and test scores the entire time-series, from $t=\SI{2}{\second}$ to $t=\SI{12}{\second}$, is evaluated, for the dyn versions, only the points in the initial two seconds ($t=\SI{2}{\second}$ to $t=\SI{4}{\second}$) are assessed. Thus, the dyn scores indicate the accuracy during the initial transients after the disturbance. Clearly, GBT exhibits the best performance with respect to the R2-score, RMSE, and MAPE. Surprisingly, the benchmark Linreg behaves second best for predictions of $i_p$, while the NN's MAPEs for $i_p$ are particularly high and close to elnet, which generally is the poorest model. For $i_q$, the RMSE and MAPE metrics deviate significantly less, while the R2-score deviates more than for $i_p$. 
\afterpage{\begin{figure}[t!]
    \centering
    \includegraphics[width=\linewidth]{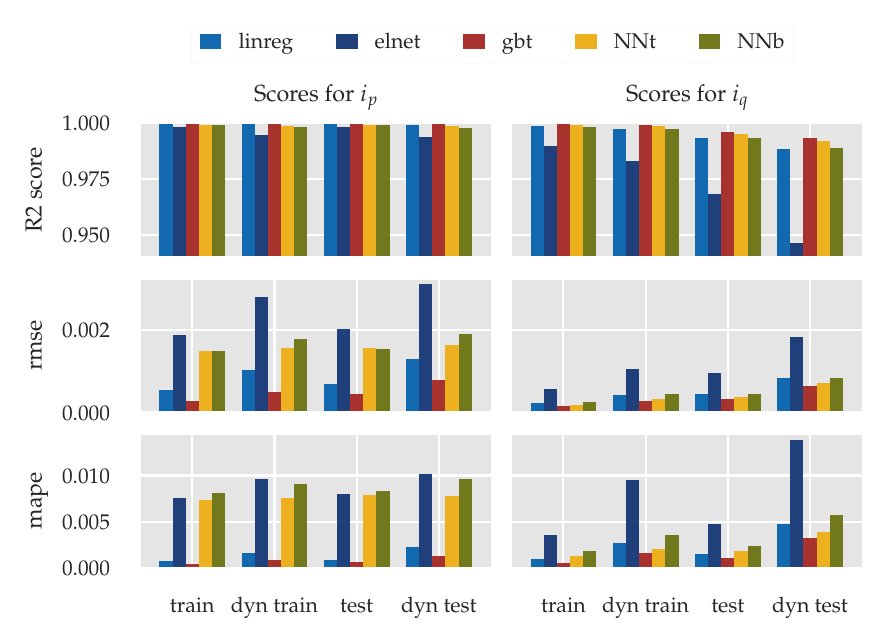}
    \caption{Training and testing scores per ML algorithm for $i_p$ in the left column and $i_q$ in the right column. While the R2 score and the mape are dimensionless, the rmse is in p.u.}
    \label{fig:scores}
\end{figure}
\begin{figure}[!t]
    \centering
    \includegraphics[width=\linewidth]{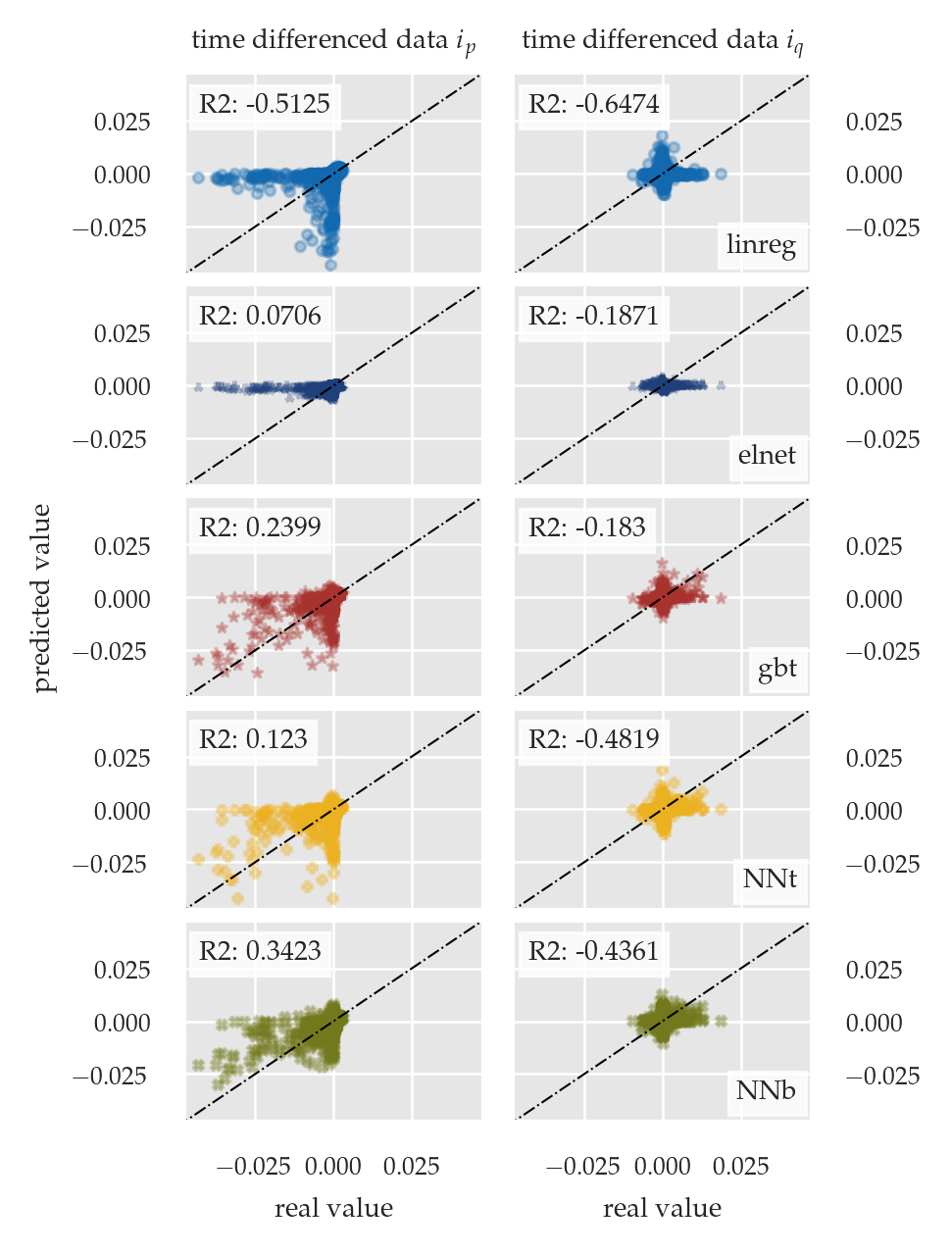}
    \caption{R2 score of the time differenced data for $i_p$ on the left and $i_q$ on the right. }
    \label{fig:dR2}
    \vspace{-10pt}
\end{figure}}

Different conclusions arise considering the scatter plots and R2-scores of the time-differenced data in Fig.~\ref{fig:dR2}. While no algorithm performs well for $i_q$, NNb performs best for $i_p$ followed by GBT, NNt, and Elnet regression. Linreg performs poorly for both currents. 

To conclude, GBT provides the best point forecast. Despite the acceptable performance of the NNs on the time-differenced data, linreg outperforms them on the other scores. Considering the application, retrieving the current shape, even with a certain offset, might be more relevant than minimizing the point-wise error without considering the evolution over time steps. Thus, time-difference metrics are potentially more important than good RMSE and MAPE scores. 

\section{Quantile Dynamic Equivalents}\label{sec:quantileforecast}
In this section, the GBT and NN techniques are implemented in a quantile fashion, allowing to predict an interval of the most probable current rather than one point for each time step. First, the required adaptions are presented. Then, evaluation metrics for quantile forecasting are introduced. An evaluation of the techniques concludes the section. 

\subsection{Quantile Forecasting}
The training process for quantile forecasting consists of minimizing the pinball loss function $J_q$, creating separate forecasts for each quantile:
\begin{align}
    J_q = \frac{1}{n_s} \sum_{i=1}^{n_s} \begin{dcases}  
        (y_i-\hat{y}_q,i) q & \text{if } y_n \leq \hat{y}_{q,i}, \\
        (y_i-\hat{y}_q,i) (1-q) & \text{if } y_n < \hat{y}_{q,i},
   \end{dcases} 
\end{align}
where $q$ is the target quantile and $\hat{y}_{q,i}$ is the prediction for quantile $q$ of the $i$-th~sample. Hence, to achieve a probabilistic forecast with nominal confidence $Q$, two quantile forecasts with $q=\pm Q/2$ are required. The loss function is asymmetric such that for any quantile higher (lower) than 50\% underestimating the target is penalized more (less) than for overestimation. Note that the 50\% quantile corresponds to the previously discussed point forecasts. In this study, quantile versions of the GBT, NNt and NNb are implemented. While quantile linear regression also exists, it was not further studied because of implementation issues in sklearn. The same features and hyper-parameters as for the point forecasts are applied. The same training and test split as before is employed, i.e. 800 time-series for training and 200 time-series for testing.  

\subsection{Quantile Evaluation Metrics}
The reliability (REL) reflects the percentage of targets that are captured within the prediction interval (PI). The average coverage error (ACE) indicates whether the reliability is in line with the PI nominal confidence. The width of the band, also called sharpness, is assessed by the average interval score (AIS). The aforementioned metrics are defined as \cite{Wan_2014}:
\begin{align}
    \text{REL}(B_Q) &= \frac{1}{n_s} \sum_{i=1}^{n_s} \mathbbm{1}_{\hat{y}_{50-Q/2}^i\leq y_i \leq \hat{y}_{50+Q/2}^i}, \\
    \text{ACE}(B_Q) &= 100\% \left(\text{REL}(B_Q)-Q\right),\\
    \text{AIS} (B_Q ) &= \frac{1}{n_s} \sum_{i=1}^{n_s} \hat{y}_{50+Q/2}^i -  \hat{y}_{50-Q/2}^i,
\end{align}
where $B_Q$ is the PI with confidence $Q$, $y_i$ is the target of sample $i$, and $\hat{y}_{50\pm Q/2}^i$ are the quantile predictions defining the lower and upper bounds of the PI. In this context, good quantile forecasts lead to an ACE close to zero and a low AIS, i.e., a good sharpness. 

\begin{figure}[!t]
\centering
\subfloat{\includegraphics[width=\linewidth]{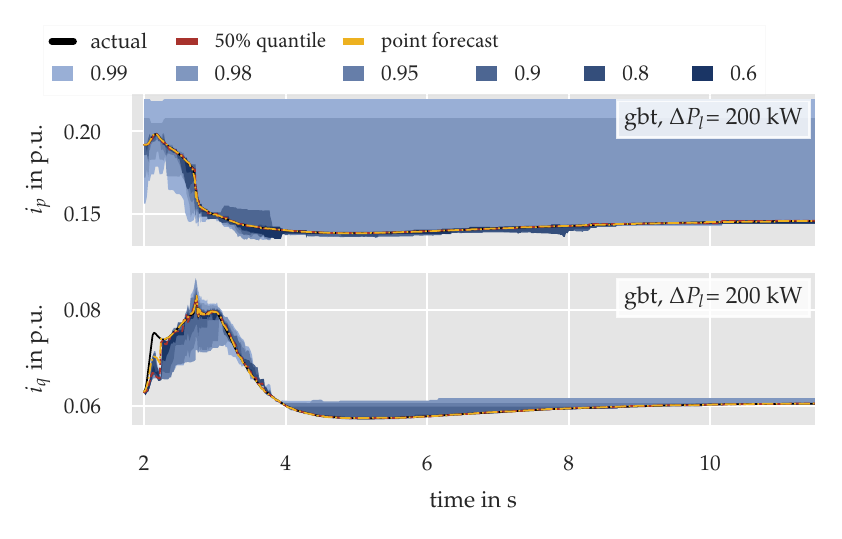}}
\vspace{-5pt}
\subfloat{\includegraphics[width=\linewidth]{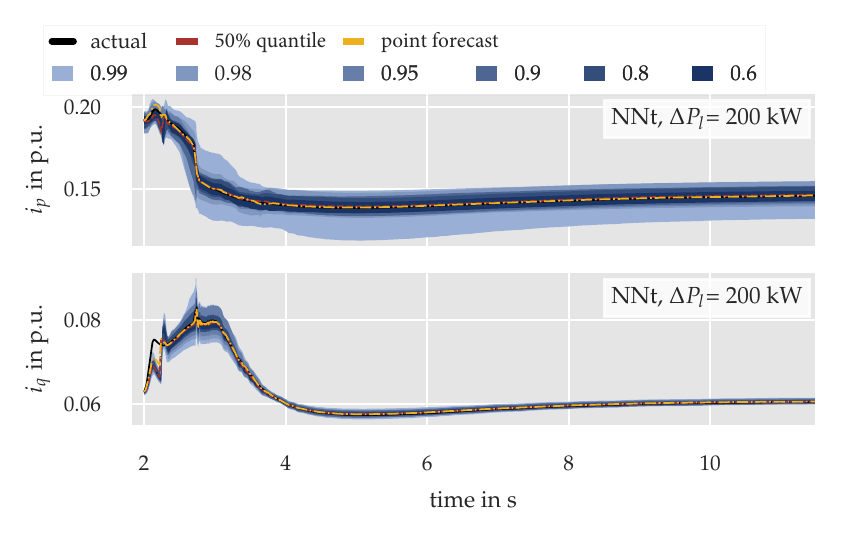}}
\vspace{-5pt}
\subfloat{\includegraphics[width=\linewidth]{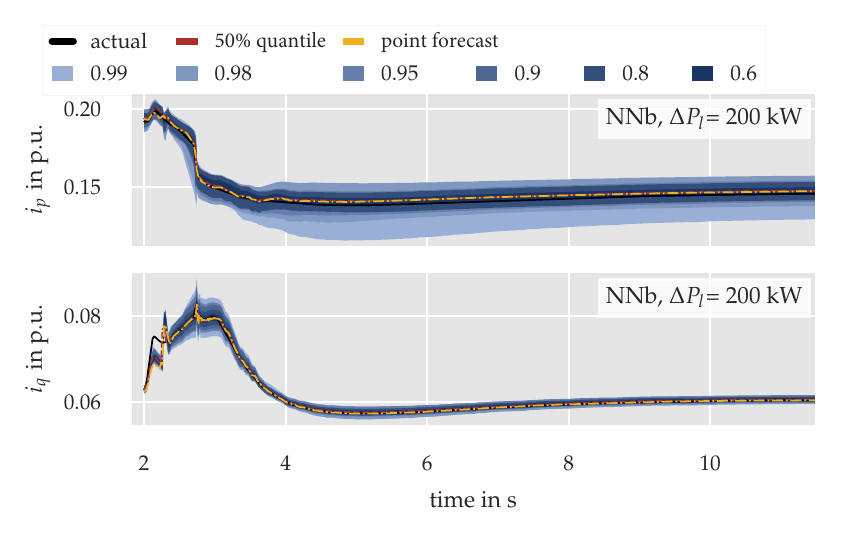}}
\caption{Time-domain performance of the quantile forecasts, GBT on the top, NNt in the middle and NNb on the bottom. A \SI{200}{\kilo\watt} increase in load is applied. }
\label{fig:TD_QuantileForecasts}
\end{figure}

\subsection{Evaluating the Quantile Current Forecasts}
Fig.~\ref{fig:TD_QuantileForecasts} shows the resulting quantile forecasts for each algorithm for an exemplary load step of \SI{200}{\kilo\watt} and different PIs. The mean response equals the point forecast from the previous section, while the median response corresponds to the $50\%$-quantile using the pinball loss function. 

The showcased quantile forecasts are relatively smooth for the two NNs, while those obtained with GBT exhibit fast-changing behavior. The performance on $i_q$ appears similar for all techniques. All fail to predict the uncertainty around the initial transient. Distinct differences between GBT and the NNs occur for $i_p$. The width of the NNs PIs seems comparable and linearly dependent on the nominal confidence provided in the legend. However, for the GBT method, the 99~\% and 98~\% confidence result in an extensive band, while the other bands appear significantly tighter than the NNs. 

Similar findings are indicated in Table~\ref{tab:QR_Scores} that lists the ACE and AIS for all algorithms and the two targets. Per definition, the width of the uncertainty band, the AIS score, increases with growing confidence interval. There is, nevertheless, a sudden increase in AIS from 95\% to 98\% confidence interval produced by the GBT for the estimation of $i_p$. The ACE scores suggest that the GBT generally underestimates the PI nominal confidence, in contrast to the NNs which tend to produce wider quantile bands. This effect diminishes when considering higher confidence intervals.

In conclusion, determining the best quantile forecasting approach depends on the application and desired properties. Concretely, a trade-off between reliability and sharpness shall be found, which also depends on the confidence interval of interest. Concerning the two NNs, NNt outperforms NNb slightly due to smaller AIS. If narrow bands are required, GBT is the better choice for confidence levels up to 95\%. However, this comes at the cost of lower reliability. 

\begin{table}[]
    \centering
    \footnotesize
    \renewcommand{\arraystretch}{1.1}
    \caption{Test scores of the different quantile forecasts.}
    \label{tab:QR_Scores}
    \begin{tabular}{p{1.2cm} p{0.5cm}  c c c c c c}
    \toprule
     \multicolumn{2}{l}{\textbf{Confidence}} & $60\%$ & $80\%$ & $90\%$ &$95\%$ & $98\%$ & $99\%$ \\[2pt] \hline
    \rowcolor{black!10} \multicolumn{8}{c}{\textbf{Scores for} $\boldsymbol{i_p}$} \\[2pt]  \hline
    \textbf{ACE } & gbt &-10.6 & -16.1 &-15.0 & -14.2 & -10.7 & -8.0  \\
    \textbf{in \%}& NNt & 38.7 & 18.9& 9.6&4.8 &1.5 & 1.0 \\
                   & NNb &-3.2 & 19.1 & 9.7 & 4.8 & 1.9 & 1.0 \\[2pt] 
    \textbf{AIS in} & gbt &0.6 & 1.3 &2.7 & 4.7 & 40.3 & 51.7  \\
    \textbf{$10^{-3}$ p.u.}& NNt & 5.7& 8.1 & 10.9 & 12.1&15.3 & 20.3 \\
                   & NNb & 4.4 & 9.7 & 10.3 & 13.1 & 17.0 & 20.8\\[2pt]  \hline
    \rowcolor{black!10} \multicolumn{8}{c}{\textbf{Scores for} $\boldsymbol{i_q}$} \\[2pt]  \hline
    \textbf{ACE } & gbt &-19.8&-23.5&-19.7&-14.6&-7.9&-5.3 \\
    \textbf{in \%}& NNt & 20.2&17.7 &7.5 &3.6 &0.8 & 0.2 \\
                   & NNb & 16.3 & 16.1 & 6.7 & 2.8 & 0.4 & -0.4 \\[2pt] 
    \textbf{AIS in} & gbt &0.1 &0.2 & 0.5&0.8&1.2&1.4 \\
    \textbf{$10^{-3}$ p.u.}& NNt &0.6 & 1.1& 1.3 & 1.8 & 2.1 & 2.3\\
                   & NNb & 0.7 & 1.2 & 1.8 & 1.9 & 2.3 & 2.4 \\ \bottomrule
    \end{tabular}
\end{table}

\section{Additional Assessments}\label{sec:weakTN}

This section presents two additional case studies that are performed with the trained equivalents. In the first case, the ML techniques are tested on a dataset generated with a weak instead of a strong TN. The second assessment evaluates how well the quantile forecasts match the variance in current trajectories obtained through the MC simulations.

\subsection{Placement in a Weaker Transmission System}
So far, the tuned DN equivalents have been tested on time-series obtained with the same TN equivalent used for generating the training series. However, in real TNs, the operating conditions, including inertia levels, might change. Thus, in this section, the trained point and quantile DN equivalents are tested on 200 time-series generated with a weaker TN. 

First, the data generation and adaptions to the model described in Section~\ref{sec:testsystem} are introduced. Then, the point and quantile forecasts results are showcased and discussed. 

\subsubsection{Data Generation for the Weak TN}
The same model and method as in Section~\ref{sec:gentimeseries} are applied to obtain time-series for a weak TN. The grid strength and inertia constant are adapted. While the system inertia is reduced from \SI{6}{\second} to \SI{1.5}{\second}, the short-circuit power is halved from $150~\text{MVA}$ to $75~\text{MVA}$.

Even though the same ten load steps ranging within {$\Delta P_l =\pm\SI{225}{\kilo\watt}$} are applied, they result in significantly higher frequency deviations. While for the strong grid, the maximum observed deviation was up to \SI{1}{\hertz}, for the weaker TN, the frequency deviations exceed \SI{1.2}{\hertz}. Thus, the following study subjects the trained ML algorithms to new conditions. 

\subsubsection{Evaluation of the Point Forecasts}
\begin{figure}[t]
    \centering
    \includegraphics{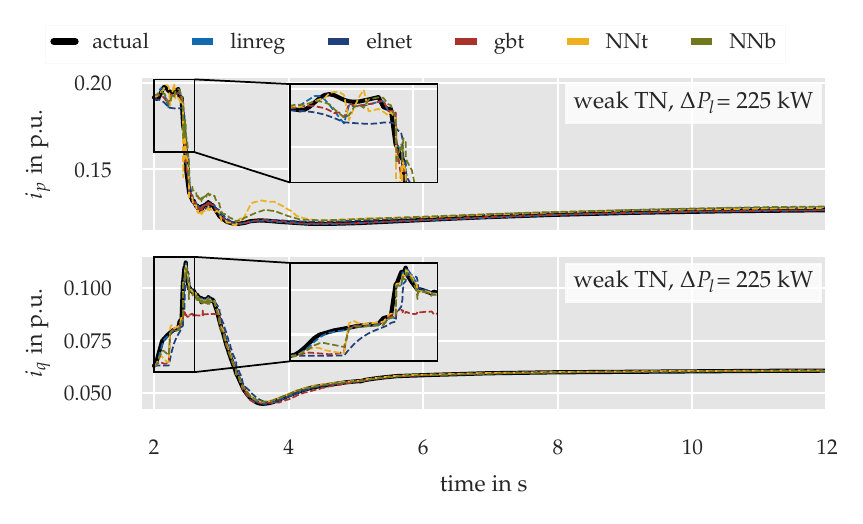}
    \caption{Time-domain performance of the point forecast on the weak TN equivalent time-series. The load increase to $\SI{225}{\kilo\watt}$ results in frequency deviations around \SI{-1.22}{\hertz}.}
    \label{fig:TD_WeakTN}
\end{figure}

The top time-domain plot in Fig.~\ref{fig:TD_WeakTN} displays an exemplary time-series for a large underfrequency event in the weak TN. The trained estimators are subject to unseen conditions, as the corresponding frequency deviations exceed \SI{1}{Hz}. Despite the untrained conditions, all of the algorithms capture the general shape of both currents. Nonetheless, significant deviations occur during the transient phase. While GBT and Linreg seem to work best for $i_p$, Linreg and NNb exhibit minor deviations from the target for $i_q$.

\begin{figure}[t]
\centering
\subfloat{\includegraphics[trim={0 6cm 0 0},clip]{Plots/AlgoComparison_Scores.pdf}
\label{fig_first_case}}
\setcounter{subfigure}{0}
\vspace{-0.5cm}

\subfloat[Strong TN equivalent]{\includegraphics[scale = 0.9, trim={0 0 1.3cm 0.6cm},clip]{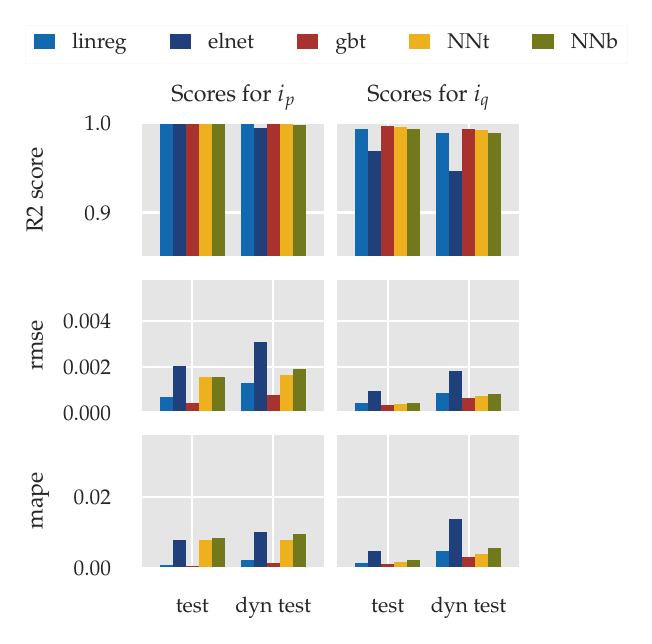}}
\subfloat[Weak TN equivalent]{\includegraphics[scale = 0.9,trim={0 0 2.5cm 0.6cm},clip]{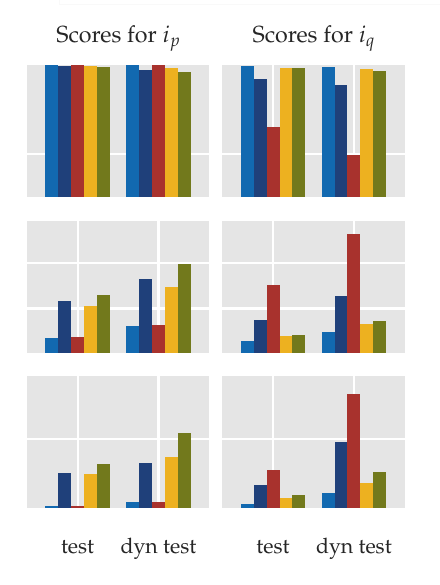}}
\caption{Comparison of the training scores for the (a) strong and (b) weak TN equivalent.}
\label{fig:weakTnscores}
\end{figure}

Fig.~\ref{fig:weakTnscores} displays the scores for each algorithm applied to the strong and the weak TN data set, and supports the previous findings. For the weak TN, Linreg and GBT have the best scores for predicting $i_p$, while GBT exhibits the worst performance on $i_q$. Linreg and Elnet scores are not significantly affected by the changing TN conditions, while all other algorithms exhibit degrading performance for the weak TN. The most significant increase in errors is observed for the NNs' prediction of $i_p$ and the GBT's estimation of $i_q$.

In conclusion, the tuned DN equivalents all forecast the general shape of the currents for the adapted TN conditions. However, Linreg appears to be the most robust.

\begin{table}[]
    \centering
    \footnotesize
    \renewcommand{\arraystretch}{1.1}
    \caption{Test scores of the different quantile forecasts when tested on the weak TN equivalent.}
    \label{tab:QR_Scores_weak}
    \begin{tabular}{p{1.2cm} p{0.5cm}  c c c c c c}
    \toprule
     \multicolumn{2}{l}{\textbf{Confidence}} & $60\%$ & $80\%$ & $90\%$ &$95\%$ & $98\%$ & $99\%$ \\[2pt] \hline
    \rowcolor{black!10} \multicolumn{8}{c}{\textbf{Scores for} $\boldsymbol{i_p}$} \\[2pt]  \hline
    \textbf{ACE } & gbt &-15.2 & -22.4&-19.2&-29.5&-18.1&-14.9 \\
    \textbf{in \%}& NNt & 34.2&14.6&8.5&4.1&1.5&0.9\\
                   & NNb &21.1&18.0&7.7&4.1&1.7&0.7 \\[2pt] 
    \textbf{AIS in} & gbt & 0.9&2.2&4.4&5.3&46.0&57.0\\
    \textbf{$10^{-3}$ p.u.}& NNt & 5.8&8.5&11.8&13.1&16.8&24.6 \\
                   & NNb & 4&10.7&10.5&13.8&17.7&22.2\\[2pt]  \hline
    \rowcolor{black!10} \multicolumn{8}{c}{\textbf{Scores for} $\boldsymbol{i_q}$} \\[2pt]  \hline
    \textbf{ACE } & gbt &-29.6&-38.3&-35.1&-29.4&-27.1&-25.6 \\
    \textbf{in \%}& NNt & 12.8&12.1&5.1&2.1&-0.7&-1.1\\
                   & NNb & 24.4&12.5&3.6&0.04&-0.8&-1.6\\[2pt] 
    \textbf{AIS in} & gbt &0.2 & 0.7 & 1.3 & 1.6 & 2.4 & 2.5\\
    \textbf{$10^{-3}$ p.u.}& NNt &0.8&1.7&1.7&2.8&2.8&3.7\\
                   & NNb & 0.9&1.7&2.6&2.7&3.0&3.3\\ \bottomrule
    \end{tabular}
\end{table}

\begin{figure}[!t]
\centering
\subfloat{\includegraphics{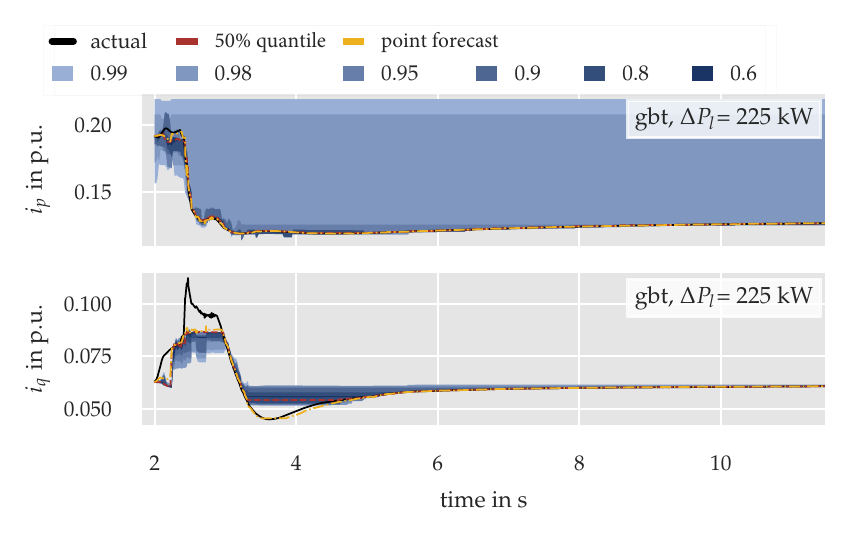}
\label{fig:QF_GBT}}
\vspace{-5pt}
\subfloat{\includegraphics{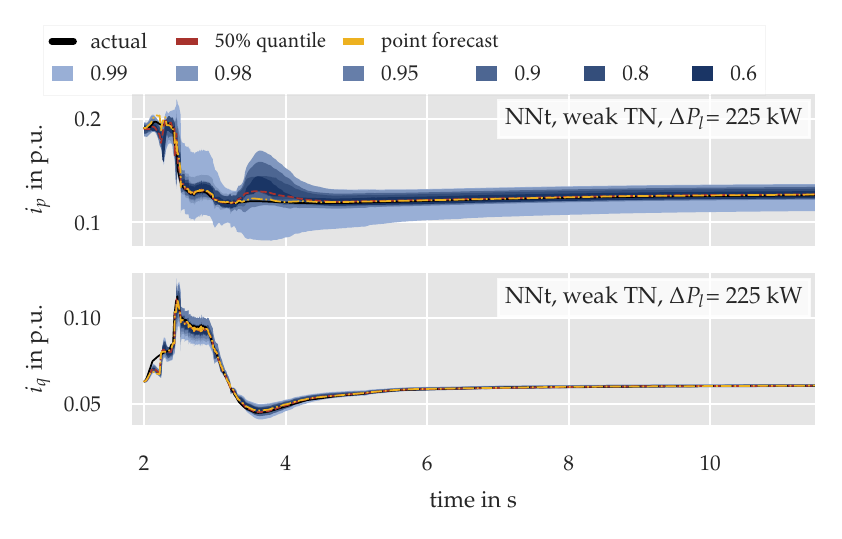}
\label{fig:QF_NNt}}
\vspace{-5pt}
\subfloat{\includegraphics{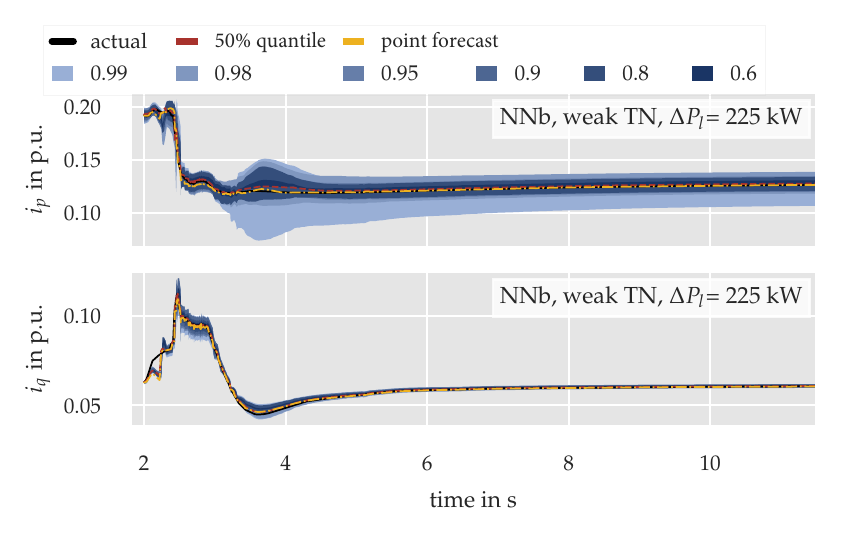}
\label{fig:QF_NNb}}
\caption{Time-domain performance of the point and quantile forecasts on the weak TN equivalent time-series. The load increase to $\SI{225}{\kilo\watt}$ results in frequency deviations around \SI{-1.22}{\hertz}.}
\label{fig:TD_QuantileForecasts_weak}
\end{figure}

\subsubsection{Evaluation of the Quantile Forecasts}
The quantile forecasts for the same \SI{225}{\kilo\watt} load increase as before is depicted in Fig.~\ref{fig:TD_QuantileForecasts_weak}. Again, the GBT prediction intervals are quite narrow for confidences up to 95\%. While the shape of $i_p$ seems to be retrieved acceptably, significant errors remain for $i_q$. The performance on $i_q$ is better for the two NNs. Similar to the strong TN case, the PIs seem to widen linearly around the mean with increasing confidence. 

Table~\ref{tab:QR_Scores_weak} supports these findings. While the GBT exhibits high AIS for high confidence intervals, it tends to predict too narrow bands, i.e. the ACE is negative. For high confidence intervals, both NNs exhibit lower AIS scores than GBT. Compared to the strong grid case the quantile bands are significantly wider, indicating that the algorithms detect the larger variety in possible current responses. 

In conclusion, all ML techniques can cope with the drastic change in TN parameters to predict $i_p$ with an increase in the bandwidths, i.e. higher AIS scores, while maintaining an adequate reliability, i.e., similar or even lower ACE scores in absolute value. Furthermore, the tuned equivalents are expected to provide a more conservative estimation of the potential current trajectories in TN studies. Nonetheless, these considerations should be proven by performing TN studies with the tuned quantile equivalents and comparing the obtained frequency and voltage trajectories to those obtained during MC simulations.

\subsection{Comparing the Quantile Equivalents to MC Simulations}
Up to this point, the quantile forecasts are assessed on specific time-series of the set of MC simulations. However, the scores and previous assessments do not reflect how well the quantile forecasts match the variance in current trajectories obtained through the MC simulations. Thus, in this qualitative assessment, the current quantile bands for one load step are compared to the MC simulation results obtained for the same step in power demand. 

\subsubsection{Method}
One random time-series from the test set is chosen for a specific load step, and the quantile forecasts are performed. The resulting bands for different confidence levels are plotted with the individual current trajectories obtained through MC simulations of the same load step. Note that the MC trajectories reflect all DN parameterizations, i.e., the training and testing sets. 

\begin{figure}[!t]
\centering
\subfloat{\includegraphics[trim = {0.9cm 0 0 0}, clip]{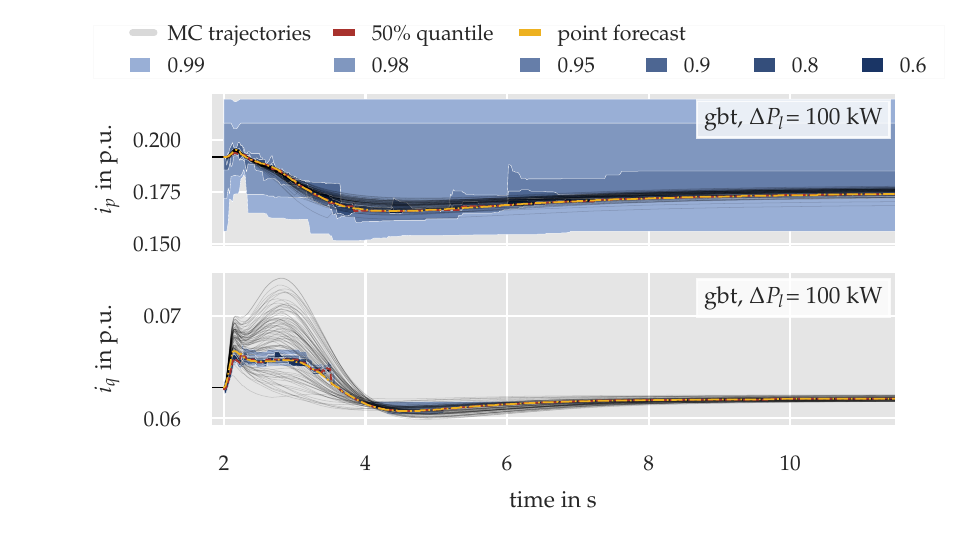}}
\vspace{-5pt}
\subfloat{\includegraphics[trim = {0.9cm 0 0 0}, clip]{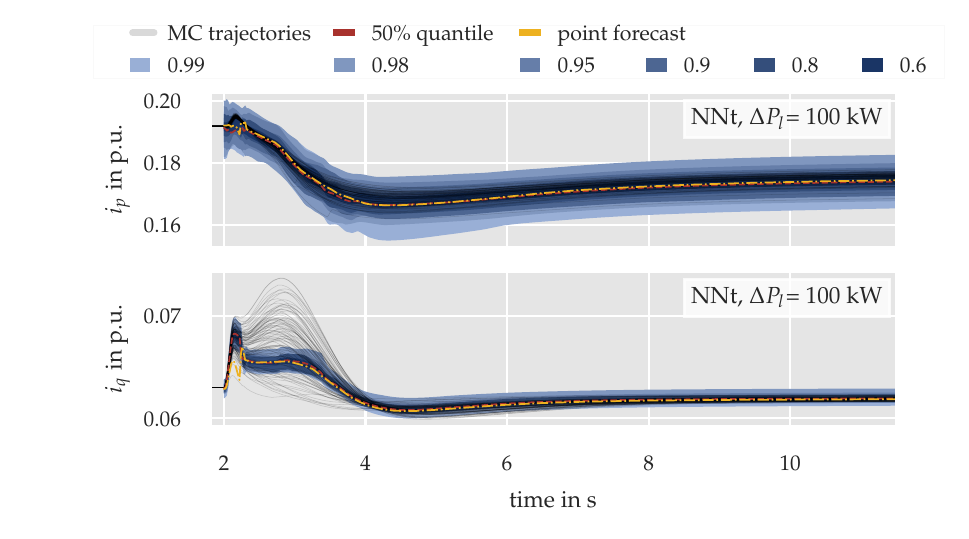}}
\caption{Time-domain performance of the quantile GBT (top) and NNt (bottom) compared against the set of all MC simulations for a step increase in load power of $\Delta P_l = \SI{100}{\kilo\watt}$. }
\label{fig:QuantileForecast_vs_MCSim}
\end{figure}

\subsubsection{Results}
Fig.~\ref{fig:QuantileForecast_vs_MCSim} displays the resulting current bands for the GBT and NNt and the MC trajectories for a step increase of \SI{100}{\kilo\watt} in load power at the TN level. The NNb is omitted as its performance is similar to the NNt.

All quantile bands for $i_p$ of the GBT and NNt overlap with the MC trajectories. However, the NNt better predicts the underlying uncertainty because the bands for high confidence intervals are narrow. In contrast, the GBT prediction intervals with a confidence of 98\% or 99\% fail to predict the current shape. Similarly, the NN provides a better estimate of the uncertainty in $i_q$. Nonetheless, the prediction interval for $i_q$ does not overlap with all MC trajectories during the initial transients. Note that these findings might differ for other step changes and for choosing another random time-series for producing the quantile forecast since it is based on one of the MC simulations.  

Despite its simplicity, the assessment highlights the potential of the proposed method for large-scale frequency stability studies. For the selected time-series, the NN can predict the uncertainty present in the currents at the TN/DN interface. Thus, the obtained current bands can increase the robustness of large-scale TN frequency stability studies, but quantitative assessment is required to support the hypotheses.  

\section{Conclusion}\label{conclusion}
This paper proposes a new approach to represent the uncertainty originating from the DN level in large-scale TN studies. Quantile forecasts are suggested to obtain the range of possible current responses at the TN/DN-interface given voltage and frequency measurements at the point of common coupling. In doing so, the confidence and robustness of large-scale TN stability studies can potentially be improved. 

For the training of several ML techniques, a rich dataset generated with MC simulations of a DN is employed. The DN contains the latest models for distributed generation and active thermal loads in low inertia system studies. 1000 time-series are produced for ten different frequency disturbances. 

In general, even the benchmark ML algorithm, linear regression, predicts the current at the TN/DN interface with acceptable accuracy. GBT provides the most accurate point forecast for active and reactive current predictions, while the NNs provide the most appropriate quantile forecasts for high confidence intervals. Case studies on a weaker TN indicate that the active current predictions are robust against changing TN conditions, while most algorithms exhibit significantly poorer performance for reactive currents. In general, the prediction intervals widen, indicating that the algorithms correctly account for the increased variety in possible current responses. 

A qualitative comparison indicates that the quantile active current bands obtained with the NNs overlap well with the current trajectories obtained through MC simulations. Nonetheless, a quantitative assessment is required to support these findings. 

Future work should consider additional disturbances, e.g., short circuits at TN level, in the training dataset to combat the poorer performance in the reactive current. Another extension of the presented work is to consider different pre-disturbance loadings of the DN.  
In addition, the tuned equivalents should be reintegrated into TN level studies to assess if the resulting frequency band equals the one from MC simulations. Two potential ways are appealing: The quantile NNs can be integrated into pyRamses by using matrix multiplications with the weights determined by the corresponding ML algorithms. On the other hand, pyRamses offers a python interface and permits interrupting time-domain simulations at a fixed rate. During each of these interruptions, the original ML algorithms in python could update the current predictions of the DN that would be kept constant until the next interruption. 

\balance
\bibliographystyle{IEEEtran}
\bibliography{bibliography}
\end{document}